# How to derive an advantage from the arbitrariness of the g-index

Michael Schreiber

*Institute of Physics, Chemnitz University of Technology, 09107 Chemnitz, Germany.*
*E-mail: schreiber@physik.tu-chemnitz.de*

The definition of the g-index is as arbitrary as that of the h-index, because the threshold number $g^2$ of citations to the *g* most cited papers can be modified by a prefactor at one's discretion, thus taking into account more or less of the highly cited publications within a dataset. In a case study I investigate the citation records of 26 physicists and show that the prefactor influences the ranking in terms of the generalized g-index less than for the generalized h-index. I propose specifically a prefactor of 2 for the g-index, because then the resulting values are of the same order of magnitude as for the common h-index. In this way one can avoid the disadvantage of the original g-index, namely that the values are usually substantially larger than for the h-index and thus the precision problem is substantially larger; while the advantages of the g-index over the h-index are kept. Like for the generalized h-index, also for the generalized g-index different prefactors might be more useful for investigations which concentrate only on top scientists with high citation frequencies or on junior researchers with small numbers of citations.

**Introduction**

On first sight, the h-index, defined as the largest number *h* of publications which have received at least *h* citation each, does not depend explicitly on any parameter. However, one can easily introduce a prefactor and require that the *h* publications have received at least $q*h$ citations each. This arbitrariness was already noted by Lehmann, Jackson, and Lautrup (2006, 2008) as well as Ellison (2010). Van Eck and Waltman (2008) utilized the prefactor to define a generalized index. I have recently analyzed the citation records of 26 physicists and showed that the prefactor can substantially influence the ranking (Schreiber, 2013).

A disadvantage of the h-index is that further citations to the publications in the h-core, i.e., the *h*-defining set of publications, do not have any effect. To overcome this disadvantage, Egghe (2006) proposed the g-index as the largest number *g* of papers which have received at least $g^2$ citations together, or equivalently at least *g* citations on average (Schreiber, 2010). Again this definition does not explicitly involve any parameter, but like the h-index, it is arbitrary in so far as one can also utilize a prefactor and demand that the *g* papers should have received at least $q*g^2$ citations together or $q*g$ citations on average. This generalization has also been proposed already by van Eck and Waltman (2008) in a slightly different way allowing for non-integer index values.

In the present paper I present a case study of the arbitrariness by investigating the generalized g-index for the citation records of 26 physicists in analogy to my previous investigation of the generalized h-index (Schreiber, 2013). Like for the h-index, the prefactor *q* influences the ranking in terms of the generalized g-index, but the effect is not so strong, because the averaging in the definition of the g-index smoothes the citation distribution curves. Consequently, a prefactor $q = 2$ leads only to very small changes in the ranking. On the other hand, for this value of $q$ the sizes of the 26 g-cores are of the same order as those of the h-cores and thus much smaller than the cores for the original g-index. Consequently the precision problem, namely to verify that all papers in the core have indeed been published by the investigated author and that no highly cited publications are missed, is substantially reduced. This means that a major disadvantage of the g-index in comparison with the determination of the h-index can be avoided without losing the advantages of the g-index.

The paper is organized as follows: In the next section the definition of the generalized g-index is given and visualized; changes in the ranking for the 26 datasets are shown and discussed. In the following section the results for the prefactor $q = 2$ are compared with the usual h-index and the A-index. In a final section further discussions and concluding remarks are given.

## The arbitrariness of the g-index

The citation records of 26 physicists from my home Institute of Physics at Chemnitz University of Technology have been obtained from the Web of Science in January and February 2007 (Schreiber, 2007) and used for an investigation of the g-index (Schreiber, 2008). All data were thoroughly checked with regard to homonyms in order to confirm the integrity of the raw data.

The determination of the g-index can be visualized by depicting the averaged citation distributions as in Fig. 1, where the average number of citations

$$\bar{c}(r) = \frac{1}{r} \sum_{i}^{r} c(i), \qquad (1)$$

is plotted versus the rank $r$ which each paper gets by sorting according to the number of citations $\bar{c}(r)$. The intersection of these histograms with the diagonal $\bar{c}(r) = r$ yields the g-indices, i.e. $\bar{c}(g) = g$. Due to the discreteness of the citation distribution, this condition is not always fulfilled and in order to be precise, one has to determine the largest value of $g$ which satisfies

$$g \leq \bar{c}(g). \qquad (2)$$

This condition implies that $\bar{c}(g+1) < g+1$.

It is equivalent to the original definition by Egghe (2006) in terms of the sum of citations

$$s(r) = \sum_{i=1}^{r} c(i) = r\ \bar{c}(r), \qquad (3)$$

namely that $g$ is the largest rank for which

$$s(g) \geq g^2. \qquad (4)$$

According to Eq. (2) the g-index gives approximately the average number of citations in the g-core. Similarly, the A-index gives the average number of citations in the h-core

$$\bar{c}(h) = A. \qquad (5)$$

I have already previously discussed the g-index for the present datasets in comparison with *h* and *A* (Schreiber, 2008).

Using a piecewise linear interpolation of the average number of citations between $\bar{c}(r)$ and $\bar{c}(r+1)$ what corresponds to an integration of the original histogram for the citation distribution *c(r)* as proposed by van Eck and Waltman (2008) one can define a continuous index $\tilde{g}$ which exactly fulfills $\bar{c}(\tilde{g}) = \tilde{g}$. A slightly different continuous version was already suggested by Rousseau (2006) and utilized in my previous comparison (Schreiber, 2008). It is visualized in Fig. 1 for two datasets, but shall not be further analyzed in the following. The usual integer results are obtained by truncating the interpolated index values, which means taking the floor function $g = \lfloor \tilde{g} \rfloor$.

The slope of the above utilized diagonal in Fig. 1 is given by $q = \tan(\alpha)$ where $\alpha = 45°$ is the angle between the diagonal and the horizontal axis, so that $q = 1$. Thus choosing a different angle $\alpha$ and in that way a different slope $q = \tan(\alpha)$ just means an arbitrary prefactor in the definition of the generalized $g_\alpha$-index, $\bar{c}(g_\alpha) = \tan(\alpha) g_\alpha = q g_\alpha$. Again, to be precise one has to search for the largest value of $g_\alpha$ which satisfies

$$q g_\alpha \leq \bar{c}(g_\alpha) \tag{6}$$

what implies $\bar{c}(g_\alpha+1) < q(g_\alpha+1)$.

In the following I visualize the influence of the proportionality factor *q* or, equivalently, of the angle $\alpha$ on the ranking. For this purpose I analyze how changing the size of the core of the most influential publications within all datasets alters the ranking. The datasets are labeled from A to Z from highest to lowest values of *h* as in the previous investigations. The averaged citation curves of datasets H, J, M, O, P, and Q are presented in Fig. 1. Obviously choosing another angle instead of 45° for the diagonal leads to different values of $g_\alpha$ and sometimes to a different ranking of the scientists.

The dependence of the indices $g_\alpha$ on *q* is shown in Fig. 2. Here the datasets have been sorted using the original g-index so that the middle line for $\alpha=45°$, i.e. $q=1$ is monotonously increasing in the plot. The angle $\alpha$ has been increased and decreased in steps of 5°. Corresponding values for *q* are denoted in Table 1. This table includes also the accumulated number $n(g_\alpha)$ of papers that belong to all the g-cores for a given $\alpha$, i.e., how many papers contribute to the 26 $g_\alpha$-indices; this is equivalent to summing the $g_\alpha$ values for each $\alpha$.

In Fig. 2 already for $\alpha = 50°$ small fluctuations can be detected which interrupt the monotony. For larger values of $\alpha$ and *q* the deviations increase and appear for different datasets, so that the ranking is changed more frequently. But in comparison with the respective erratic changes of $h_\alpha$ which were presented in my previous investigation (Schreiber, 2013) the fluctuations of $g_\alpha$ are relatively small. This is not surprising, because the averaging of the citation frequencies leads to rather smooth distribution functions of $\bar{c}(r)$ in comparison with *c(r)*. These smoother functions appear in Fig. 1 already, showing smaller steps than the corresponding original citation distributions.

The deviations are stronger for smaller values of $q$ and α decreasing towards 0, where the lowly cited publications in the long tail of the citation distributions become relevant, too. But also in this range due to the averaging the fluctuations are relatively small compared to the respective behavior of the $h_\alpha$-index.

The changes in the ranking are visualized in Fig. 3. Some ranks fluctuate strongly for large α. Note that in this range only few papers contribute and thus the smoothing effect due to the averaging is small, because not so many citation frequencies are averaged. Otherwise the changes are small and usually monotonous. At the top of the ranking nothing is altered, and at the bottom small changes lead to tied ranks and sometimes to interchanged ranks. A noteworthy exception is the dataset G which rises from rank 12 for $g_{5°}$ to 9 before dropping to 13.5 and ending again at rank 12 for $g_{85°}$. The strongest drop occurs for scientist N from 11 to 18.5, the largest improvement for researcher X from 23 to 11 monotonously except for one instance. Similarly scientist P advances from rank 13.5 for $g_{5°}$ to rank 5 for $g_{85°}$ also monotonously except for one instance. On the other hand researcher H remains at ranks 7, 7.5, or 8 for α = 5° up to α = 80° and only then drops fast to rank 12.0.

In order to quantify these changes, Pearson's correlation coefficients κ and Spearman's rank-order correlation coefficients $κ_s$ between the original g-index and the $g_\alpha$-values are presented in Table 1. In principle, when comparing rank orders, one should use Spearman's correlation coefficients. However, as there are several studies which have utilized Pearson analysis, I also show the respective values in Table 1.

For the Pearson analysis the correlations are very strong, with correlation coefficients close to 1 and smaller than 0.985 only for the extreme cases of α = 5° and α = 85°. The rank-order correlation coefficients are slightly smaller, because small differences in the index values can lead to larger deviations of the ranks especially if several datasets are tied, i.e. have identical index values and thus the same rank. But in the present case even these coefficients remain larger than 0.97 except for α = 5° and α = 85°.

In conclusion, exploiting the arbitrariness of the g-index leads to small changes in the ranking. In order to test whether the large correlation coefficients might be caused by the relatively stable ranks of the top and bottom datasets, I have repeated the analysis excluding the top 3 and bottom 3 datasets. The calculated correlation coefficients for this subset are smaller, see Table 1, but still above 0.94 except for α = 5° and α = 85°.

**Selecting one of the generalized g-indices**

The total number of papers in all the g-cores for α = 45° is $n(g_{45°}) = 623$, see Table 1. This is much larger than the accumulated number of papers in the h-cores which comprise 387 papers. As a consequence the determination of the original g-index is much more involved than the calculation of the h-index, because many more publications have to be checked with respect to the question whether they have really been published by the investigated author. In my view, this enlarged precision problem is the major disadvantage of the g-index. Choosing a prefactor $q > 1$ of course reduces the size of the cores. In particular, for α = 65° one finds $n(g_{65°}) = 380$ so that it is tempting to use this value of α and replace the original g-index by $g_{65°}$.

However, α = 65° corresponds to a proportionality factor $q = 2.14$ which is somewhat awkward. Therefore, I propose to utilize the prefactor $q = 2$, corresponding to α = 63.4°, see Fig. 1. The total number of papers in the g-cores then amounts to $n(g_{63.4°}) = 395$. In order to simplify the notation in the following I shall utilize $q$ instead of α as the characterizing subscript for the generalized g-index, so that $g_{63.4°} \equiv g_2$. The resulting values are presented in Table 2 in comparison with the values of the original g-index and the h-index. The ratio between $g_2$ and $g$ ranges from 0.56 (researcher Y) to 0.68 (researchers E and I), on average it is 0.63±0.03. The small standard deviation suggests that the changes in the ranking are small, this is verified by the values in Table 2: The ranks in terms of $g$ and $g_2$ differ by at most one position, except for researcher O. But this is an unusual case which can be related to tied ranks: O, R and S, X are tied for $g$, while R, X and O, S (as well as Q, V) are tied for $g_2$. It is interesting to note, that although the prefactor leads to much smaller values of the $g_2$-index, the number of tied papers does not increase: There are 16 papers in 8 ties for $g$, and 16 papers in 7 ties for $g_2$. In comparison, there are 14 papers in 5 ties for $h$.

The comparison between $h$ and $g_2$ shows larger differences. The ratio between $g_2$ and h ranges from 0.69 (Q) to 1.50 (X), on average it is 1.02±0.18. In correspondence with the relatively large standard deviation, the rankings differ substantially, e.g. for P and Q the same value $h = 13$ puts them at rank 16, but the different results of $g_2 = 16$ and 9 correspond to ranks 8.5 and 22.5, respectively. This reflects the strongly different citation distributions: the $g_α$-index awards non-homogeneous citation impact of the highly-cited publications.

This is of course also true for the A-index. In fact, the effect is even stronger, as can be seen from Table 2, where the $g_2$-values are between $h$ and $A$ in most cases. The exceptions can be traced back to the discreteness of the h- and the $g_2$-index and would not appear, if the continuous versions were used. This can be seen from the definitions, because $h < \tilde{g}_2$ implies $\bar{c}(h) > \bar{c}(\tilde{g}_2) = 2\tilde{g}_2$ and therefore $A/2 > \tilde{g}_2 > h$. In this way the A-index counterbalances the disadvantage of the h-index, namely that the more-than-$h$ citations to the publications in the h-core are not taken into account. So in principle, one does not need the $g_2$-index. However, utilizing the A-index means that one has first to determine the h-index and thus one ends up with two characteristic values for each researcher. The question then is, whether this is helpful. In my view, the g-index or rather the $g_2$-index is sufficient.

The relatively small differences between the indices can be quantified: Pearson's correlation coefficients κ and Spearman's rank-order correlation coefficients $κ_s$ which are given in Table 3, are all very close to 1. The very large values for the comparison of $g$ and $g_2$ are not surprising in view of the large values in Table 1. But also the comparison with $h$ and $A$ yields values larger than 0.95 for the Pearson and larger than 0.91 for the Spearman analysis. Only the correlation between $h$ and $A$ is not quite so strong. But I do not think, that this justifies the need for both indices. In any case, all these indices are quite noisy indicators if one wants to employ them for measuring the scientific impact of the publication record of a researcher. Due to this intrinsic uncertainty the small differences should not be utilized to value one researcher better or worse than the other.

This leads to the question, which of the discussed indices should be discarded. As already mentioned, $A$ cannot be kept alone, because it depends on $h$. Because of its simplicity, one might want to favor $h$ instead of $g$ and $g_2$. However, it appears somewhat unfair that more than $h$ citations to the papers in the h-core do not have an effect. Therefore, from this point of view $h$ should be discarded. Due to the apparently unnecessarily large g-cores, $g$ should be discarded in favor of $g_{63.4°}$ from a practical point of view.

Whether the proportionality factor $q = 2$ is the best choice remains an open question. I have previously argued for the generalized h-index (Schreiber, 2013) that very large values of $q$ should not be used, because then too few highly cited papers are taken into account. On the other hand very small values of $q$ emphasize the long tail of lowly cited publications and are therefore also not suitable.

Ravallion and Wagstaff (2011) have critizised the missing theoretical foundation of the h-index as well as the g-index. They propose to determine the scientific impact or the scholarly influence by defining an influence function which is then evaluated for the complete citation profile yielding the qi-index. Obviously this aggravates the precision problem, because all publications of a scientist are taken into account. While Ravallion et al. (2011) agree that additional citations should always increase the influence functions, they question the g-index, because it gives the same weight to all additional citations to publications in the g-core and no weight to the lowly-cited papers. Rather they favor diminishing marginal influence so that the first citation to a given publication has the highest impact, while a further citation to an already highly-cited paper is considered less significant and is attributed a smaller influence; which is even vanishing if the most cited paper is concerned. This is a plausible concept and shows that it is quite subjective, whether one prefers one or another variant of the bibliometric measures. However, due to practical considerations it is rather unlikely that the much more involved determination of the qi-index will be performed in large evaluations, in spite of the sound mathematical foundation on which it is based.

**Conclusion**

One cannot define the same best value of $q$ for all evaluations. It is convenient to use the standard $q = 1$. But I have already suggested (Schreiber, 2013) that values of $q > 1$ might be more practical for top scientists with a much more skewed citation record. In particular, for the generalized h-index I have proposed to utilize a proportionality factor of $q = 3$ for comparing eminent scientists, because such a relatively large $q$ value would reasonably reduce the size of their h-cores which is unnecessarily large for the original definition with $q = 1$. For the generalized $g_\alpha$-index this would correspond to a prefactor of $q = 5.4$ if the objective is again to get about the same accumulated core sizes. For a group of more average scientists like in the present study, in comparison with the original h-index the above discussed prefactor $q = 2$ for the generalized g-index appears to be reasonable. For an evaluation of a group of junior scientists I have proposed to use a value of 1/3 as proportionality factor of the generalized h-index in order to achieve a better distinguishability by increasing the size of the h-cores. Judging from the accumulated number of papers in the 26 cores this would correspond to $q = 0.9$ in the case of the generalized index $g_\alpha$.

In conclusion, now I recommend to use the generalized index $g_\alpha$ instead of the h-index, the A-index, and also instead of the original g-index. For top scientists with high citation numbers I suggest to utilize the prefactor $q = 5$, for more average scientists $q = 2$, and for junior people $q = 1$.

TABLE 1. Pearson's correlation coefficients $\kappa$ and Spearman's rank-order correlation coefficients $\kappa_s$ between the original g-index and the $g_\alpha$-values for the data shown in Figs. 2 and 3; the correlation coefficients after excluding the top 3 and the bottom 3 datasets from the sample are also given; all correlation coefficients are significant at the 0.001 level as determined by the t distribution. $q$ is the slope of the $g_\alpha$-determining line (compare Fig. 1) and $n(g_\alpha)$ is the accumulated number of papers in all g-cores for each value of $\alpha$.

| $\alpha$ | 5° | 10° | 15° | 20° | 25° | 30° | 35° | 40° | 45° |
|---|---|---|---|---|---|---|---|---|---|
| $q=\tan\alpha$ | 0.09 | 0.18 | 0.27 | 0.36 | 0.47 | 0.58 | 0.70 | 0.84 | 1.00 |
| $n(g_\alpha)$ | 2345 | 1691 | 1350 | 1135 | 979 | 869 | 774 | 692 | 623 |
| $\kappa$(A-Z) | 0.959 | 0.985 | 0.990 | 0.994 | 0.996 | 0.997 | 0.999 | 0.999 | 1.000 |
| $\kappa_s$(A-Z) | 0.959 | 0.970 | 0.976 | 0.982 | 0.987 | 0.994 | 0.997 | 0.996 | 1.000 |
| $\kappa_s$(C,D,F-W,X) | 0.915 | 0.940 | 0.951 | 0.963 | 0.974 | 0.989 | 0.995 | 0.992 | 1.000 |

| $\alpha$ | 50° | 55° | 60° | 65° | 70° | 75° | 80° | 85° |
|---|---|---|---|---|---|---|---|---|
| $q=\tan\alpha$ | 1.19 | 1.43 | 1.73 | 2.14 | 2.75 | 3.73 | 5.67 | 11.43 |
| $n(g_\alpha)$ | 555 | 496 | 439 | 380 | 322 | 257 | 192 | 113 |
| $\kappa$(A-Z) | 0.999 | 0.999 | 0.998 | 0.996 | 0.994 | 0.993 | 0.986 | 0.967 |
| $\kappa_s$(A-Z) | 0.997 | 0.997 | 0.995 | 0.992 | 0.987 | 0.985 | 0.973 | 0.933 |
| $\kappa_s$(C,D,F-W,X) | 0.994 | 0.993 | 0.989 | 0.983 | 0.971 | 0.968 | 0.945 | 0.854 |

TABLE 2. Values of the h-index, the g-index, the A-index, and the $g_\alpha$-index for the 26 datasets, where a prefactor $q = 2$ corresponding to $\alpha = 63.4°$ is utilized ($g_{63.4°} \equiv g_2$); the datasets are sorted using the $g_\alpha$-index; the rank order for the indices is given by $\mathcal{O}(\text{index})$. For tied papers the average rank is calculated.

| Dataset | h | g | $g_2$ | A | $\mathcal{O}(h)$ | $\mathcal{O}(g)$ | $\mathcal{O}(g_2)$ | $\mathcal{O}(A)$ |
|---|---|---|---|---|---|---|---|---|
| A | 39 | 67 | 43 | 93.9 | 1.0 | 1.0 | 1.0 | 1.0 |
| B | 27 | 45 | 29 | 62.6 | 2.0 | 2.0 | 2.0 | 2.0 |
| E | 19 | 37 | 25 | 62.4 | 5.0 | 3.0 | 3.0 | 3.0 |
| C | 23 | 36 | 23 | 47.3 | 3.0 | 4.0 | 4.0 | 4.0 |
| I | 15 | 28 | 19 | 46.1 | 9.5 | 6.0 | 5.0 | 5.0 |
| D | 20 | 29 | 18 | 35.5 | 4.0 | 5.0 | 6.0 | 8.0 |
| H | 16 | 26 | 17 | 35.9 | 8.0 | 7.5 | 7.0 | 7.0 |
| F | 18 | 26 | 16 | 32.2 | 6.0 | 7.5 | 8.5 | 11.0 |
| P | 13 | 24 | 16 | 41.5 | 16.0 | 9.5 | 8.5 | 6.0 |
| J | 15 | 23 | 15 | 32.1 | 9.5 | 11.5 | 10.5 | 12.0 |
| M | 14 | 24 | 15 | 34.0 | 12.5 | 9.5 | 10.5 | 10.0 |
| G | 17 | 23 | 14 | 28.4 | 7.0 | 11.5 | 12.5 | 14.0 |
| L | 14 | 22 | 14 | 30.6 | 12.5 | 13.5 | 12.5 | 13.0 |
| K | 14 | 21 | 13 | 27.7 | 12.5 | 15.0 | 14.5 | 15.5 |
| N | 14 | 22 | 13 | 27.7 | 12.5 | 13.5 | 14.5 | 15.5 |
| R | 12 | 19 | 12 | 27.0 | 18.5 | 16.5 | 16.5 | 17.0 |
| X | 8 | 18 | 12 | 35.1 | 24.0 | 18.5 | 16.5 | 9.0 |
| O | 13 | 19 | 11 | 22.8 | 16.0 | 16.5 | 19.5 | 20.0 |
| S | 12 | 18 | 11 | 22.8 | 18.5 | 18.5 | 19.5 | 21.0 |
| U | 10 | 17 | 11 | 23.7 | 21.0 | 20.5 | 19.5 | 19.0 |
| V | 10 | 17 | 11 | 24.4 | 21.0 | 20.5 | 19.5 | 18.0 |
| Q | 13 | 15 | 9 | 17.1 | 16.0 | 22.5 | 22.5 | 23.0 |
| T | 10 | 15 | 9 | 18.0 | 21.0 | 22.5 | 22.5 | 22.0 |
| W | 9 | 13 | 8 | 15.6 | 23.0 | 24.0 | 24.0 | 25.0 |
| Z | 5 | 10 | 6 | 17.0 | 26.0 | 25.0 | 25.0 | 24.0 |
| Y | 7 | 9 | 5 | 11.0 | 25.0 | 26.0 | 26.0 | 26.0 |

TABLE 3. Correlation coefficients between $h$, $g$, $g_2$, $A$. Values for Spearman's rank order correlation coefficients are given in the upper right triangle, values for Pearson's correlation coefficients are presented in the lower left triangle.

| | h | g | $g_2$ | A |
|---|---|---|---|---|
| h | 1.000 | 0.936 | 0.910 | 0.805 |
| g | 0.971 | 1.000 | 0.991 | 0.943 |
| $g_2$ | 0.956 | 0.997 | 1.000 | 0.970 |
| A | 0.890 | 0.971 | 0.983 | 1.000 |

FIG 1. Averaged citation distributions of 6 scientists, i.e., average citation frequencies versus paper number after sorting the papers according to the number of citations; the 5 dashed lines are plotted at angles of 75°, 60°, 45°, 30°, 15° with the horizontal axis. The dotted line reflects the prefactor $q = 2$, corresponding to the angle 63.4°. The sequence HPMJOQ of the citation curves denoted in the inset was determined (from top to bottom) in the range between the diagonal and the dotted line. In order to facilitate the distinction of the various curves for two datasets (H and M) instead of the histograms the piecewice linear interpolated functions are plotted.

FIG 2. Dependence of the generalized indices $g_\alpha$ on $q = \tan(\alpha)$ for the 26 datasets in the present investigation. $q$ increases from $q = 0.09$ (top) to 11.43 (bottom). Note that the sequence of the datasets as denoted on the horizontal axis deviates from the alphabetic order, because the datasets have been sorted according to the original g-index while the alphabetic labels are the same as in my previous investigations (Schreiber, 2008, 2013), i.e. according to the original h-index.

FIG 3. Ranking of the 26 scientists in dependence on $\alpha$, determined from the $g_\alpha$ values in Fig. 2; for tied ranks the average is given; various segments for some curves are slightly shifted by ±0.1 or ±0.2 in order to facilitate distinguishing piled-up segments; thicker lines are used for datasets which are mentioned in the text. These datasets are marked with filled symbols, the other datasets are marked with line symbols.

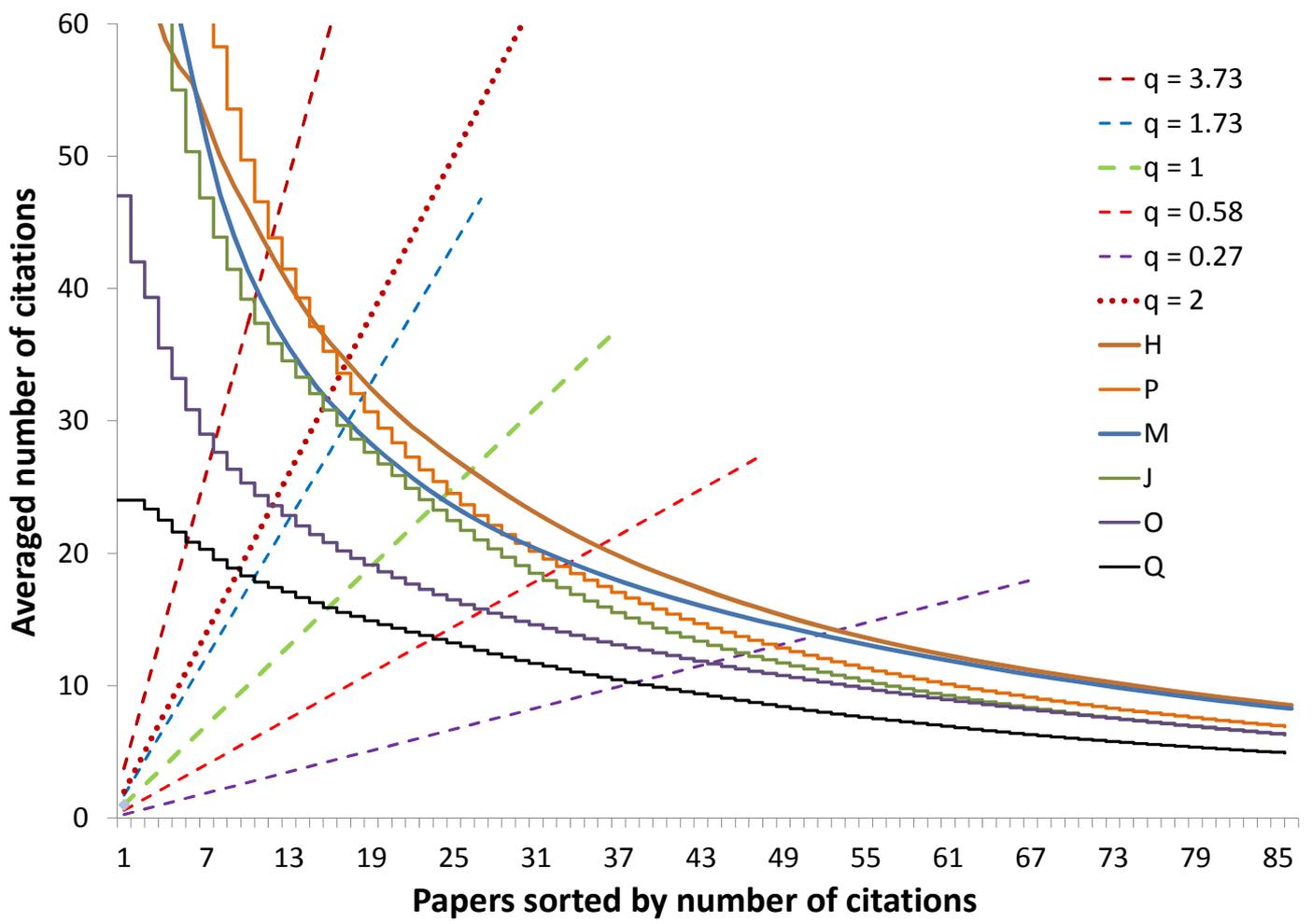

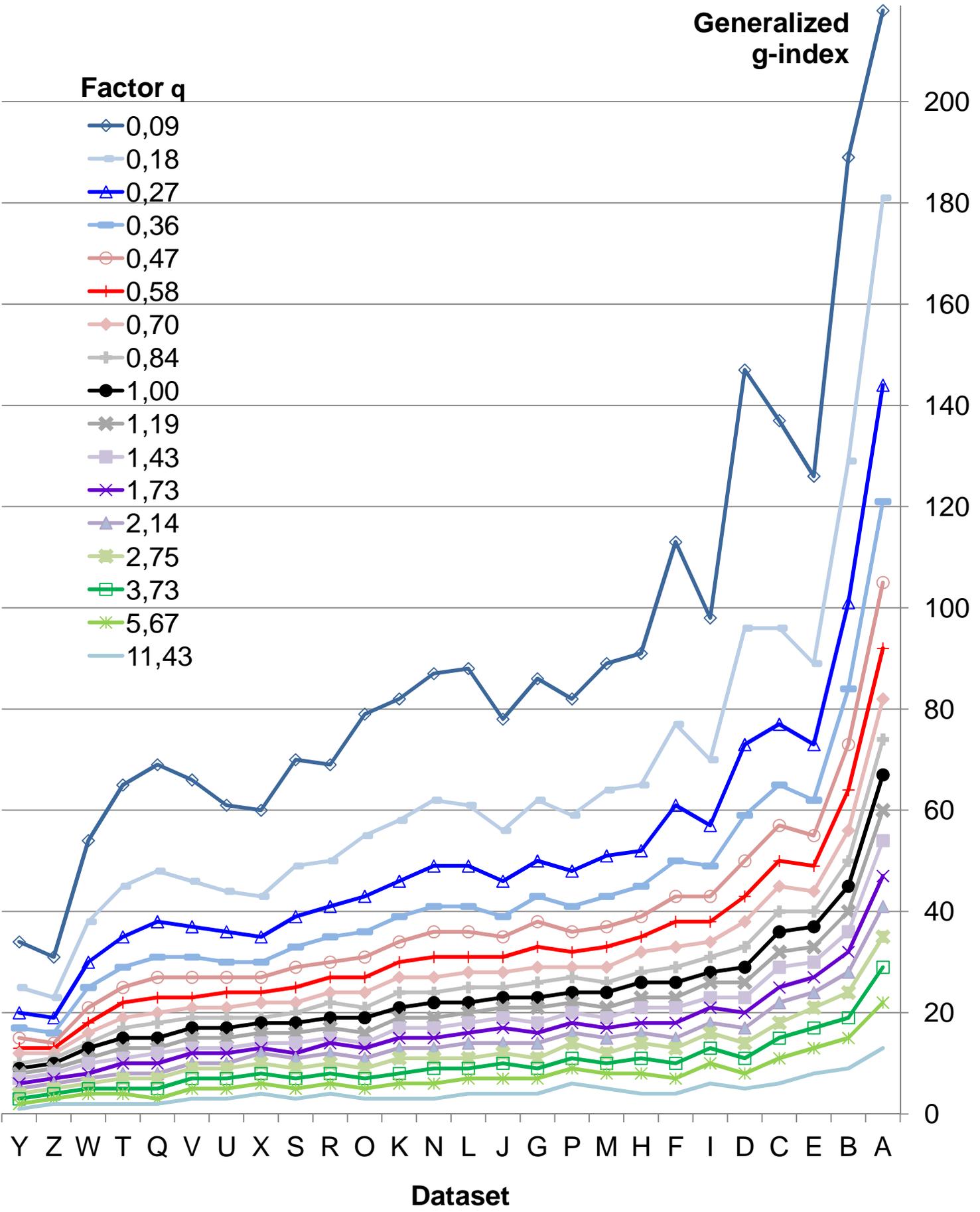